\pdfoutput=1
\documentclass{article}
\usepackage{setspace}
\usepackage{verbatim}
\usepackage{ifthen}
\newboolean{figatend}
\setboolean{figatend}{false}
\newboolean{figs}
\setboolean{figs}{true}
\ifthenelse{\boolean{figatend}}
{
  \usepackage[nofiglist,nomarkers]{endfloat}

}{}
\usepackage[margin=1in]{geometry}
\usepackage{graphics}
\usepackage[super,sort&compress]{natbib}
\begin{document}
\newcommand{\CFF}{\ensuremath{^4\mathrm{CF}}}
\newcommand{\CFT}{\ensuremath{^2\mathrm{CF}}}
\newcommand{\CFO}{\ensuremath{^1\mathrm{C}}}
\newlength{\scolwid}
\newlength{\dcolwid}
\setlength{\scolwid}{89mm} 
\setlength{\dcolwid}{183mm} 
\setlength{\scolwid}{6.5in} 
\setlength{\dcolwid}{6.5in} 
\author {O. E. Dial\footnote{Massachusetts Institute of Technology, Cambridge, MA 02139, USA}, R. C. Ashoori\footnotemark[1], L. N. Pfeiffer\footnote{Alcatel-Lucent Bell Laboratories, Murray Hill, New Jersey 07974, USA (present address: Princeton, Princeton New Jersey, 08544, USA.}, K. W. West\footnotemark[2]}
\title{Anomalous structure in the single particle spectrum of the fractional quantum Hall effect}
\maketitle
\begin{abstract}
The two-dimensional electron system (2DES) is a unique laboratory for
the physics of interacting particles.  Application of a large magnetic
field produces massively degenerate quantum levels known as Landau
levels.  Within a Landau level the kinetic energy of the electrons is suppressed, and
electron-electron interactions set the only energy scale\cite{Jain}.
Coulomb interactions break the degeneracy of the Landau levels and can
cause the electrons to order into complex ground states.  In the high
energy single particle spectrum of this system, we observe salient and
unexpected structure that extends across a wide range of Landau level
filling fractions.  The structure appears only when the 2DES is cooled
to very low temperature, indicating that it arises from delicate
ground state correlations.  We characterize this structure by its
evolution with changing electron density and applied magnetic field.
We present two possible models for understanding these observations.
Some of the energies of the features agree qualitatively with what
might be expected for composite Fermions, which have proven effective
for interpreting other experiments in this regime.  At the same time,
a simple model with electrons localized on ordered lattice sites also
generates structure similar to those observed in the experiment.
Neither of these models alone is sufficient to explain the
observations across the entire range of densities measured.  The
discovery of this unexpected prominent structure in the single
particle spectrum of an otherwise thoroughly studied system suggests
that there exist core features of the 2DES that have yet to be
understood.

\end{abstract}

Our measurements are performed using time domain capacitance
spectroscopy (TDCS)\cite{Chan97,Dial07}, which allows measurement of
the single particle density of states (SPDOS) with accurately
calibrated energy and density scales. The technique uses a repeated
series of electronic pulses to measure I/V tunnelling characteristics,
with delays between the pulses to allow the sample to
re-equilibrate. Two experimental enhancements make the results
described in this letter possible.  First, compared to our previous
work, we use a sample with a wider, 230\AA\ quantum well to confine
the electrons in two-dimensions.  This reduces the scattering from
fluctuations in the well thickness\cite{Luhman07}.  Secondly, we have
enhanced our experimental technique to allow us to measure spectra at
higher magnetic fields (see supplement).  The spacing between Landau
level orbit centers is smaller at higher fields, increasing the energy
scale of the Coulomb interaction and separating features more clearly
in the spectrum.

\begin{figure}
\ifthenelse{\boolean{figs}}{\resizebox{\scolwid}{!}{\includegraphics{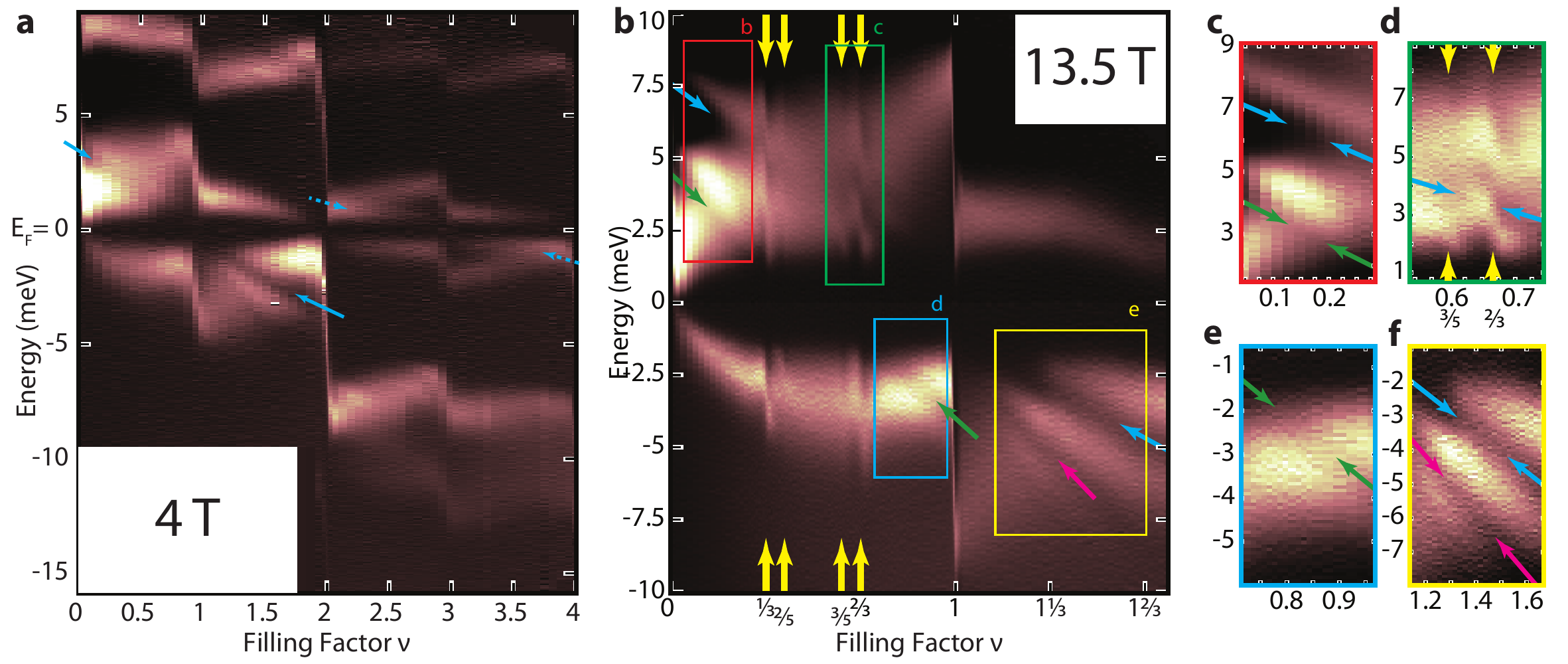}}}{}
\caption{{\bf High field TDCS spectra show ``sash'' features: bright
  and dark diagonal lines across the spectrum.} The horizontal axis in
  each spectrum is the electron density in the quantum well, expressed
  as a filling factor $\nu$ with $\nu=1$ corresponding to completely
  filling the lowest spin-polarized Landau level.  The vertical axis
  is energy measured from $E_f$, with $E>0$ corresponding to injecting
  electrons into empty states in the quantum well and $E<0$
  corresponding to ejecting electrons from filled states.  Bright
  regions correspond to high SPDOS.  In the spectrum taken at a temperature
  of 100 mK with a 4 Tesla perpendicular magnetic field
  shown in \textbf{a}, features associated with the $\nu=1$ sash
  are highlighted with a blue arrow, while another sash about $\nu=3$
  is indicated with a dashed blue arrow.  In the 13.5 Tesla data in
  \textbf{b}, taken at 80 mK, the $\nu=1$ sash is similarly indicated
  with blue arrows, while the $\nu=1/2$ sash about $\nu=1/2$ is
  indicated with green.  Sharp downward steps in the spectrum
  corresponding to chemical potential jumps at filling fractions
  corresponding to fractional quantum Hall plateaus are indicated with
  yellow arrows.  The contrast for positive and negative energies has
  been adjusted separately in this spectrum.  \textbf{c-f}, selected
  regions of the spectrum are blown up and their contrast enhanced to
  ease identification of the features, indicated with arrows that
  match the colors in \textbf{b}. }
\label{data_figure}        
\end{figure}

As described in our previous work\cite{Dial07}, as we vary the filling
factor υ from $\nu=0$ to $1$ the inter-electron spacing decreases.
This causes an increased Coulomb interaction that opens an exchange
gap between spin-up and spin-down states at the Fermi
surface. Generically, we then expect to see one peak above the Fermi
energy $E_f$ ($E=0$ in our plots), and one peak below $E_f$ in the
SPDOS, and those peaks should move away from each-other as we raise
the density.  Similarly, we expect to see exchange gaps collapse as we
raise the density in the $1<\nu<2$ range, so the two peaks will
approach $E_f$ again as we raise the density to $\nu=2$.  At higher
magnetic fields, several new features become apparent that are not
explained within this simple exchange-based framework.

At a perpendicular magnetic field $B$ of 4 Tesla, an additional
tunneling feature (the edge of this feature closest to $E_f$ is
delineated with blue arrows in Figure 1\textbf{a}) appears at roughly
4 meV near zero density, with an energy that decreases towards $E=0$
as it approaches $\nu=1$.  This trend with density is surprising; this
is a region where we expect to see the exchange gap opening, and
states correspondingly moving away from the Fermi energy.  A second
feature that appears to be a continuation of this line extends towards
increasingly negative energies between $\nu=1$ and $2$, once again
moving counter to the behaviour of the exchange gap.  We will refer to
these features as the ``$\nu=1$ sash''.  Near $E=0$, we see a dark
band of suppressed tunnelling caused by the magnetic field induced
Coulomb gap\cite{Ashoori90,Eisenstein92,Ashoori93,Yang93,Chan97}.

On increasing $B$ to 13.5 Tesla (Figure 1\textbf{b-f}), we find this
$\nu=1$ sash (blue arrows) becomes more distinct and shifts to higher
energies.  In addition, we see ``jumps'' in our data at $\nu=1/3$,
$2/5$, $3/5$, and $2/3$ (yellow arrows).  These result from chemical
potential variations at the fractional quantum Hall (FQH) plateaus, as
observed elsewhere\cite{Eisenstein94,Khrapai07}.  A second
``$\nu=1/2$'' sash also becomes visible at smaller energies,
extrapolating to zero energy at $\nu=1/2$ (green arrows).  An
additional sash also appears below the $\nu=1$ sash, in the vicinity
of $1<\nu<2$ (purple arrow).

These sashes are destroyed at 13.5 Tesla when we raise the sample
temperature to 4 K, while the exchange splitting and integer quantum
Hall features survive (see supplemental figure 1).  Despite their high
energies in the spectrum (up to 8 meV), a relatively small thermal
energy of 0.4 meV suffices to eliminate these sashes.  This
demonstrates a remarkable property of the single particle spectrum:
the high energy spectral features depend on fragile properties of the
2DES that only develop at low temperatures.

All of these sashes have the opposite density dependence than that
expected from the opening and closing of the exchange gap, show a
linear dependence in energy on filling fraction which becomes steeper
as $B$ is raised, and appear to cross the Fermi energy close to either
$\nu=1$ or $\nu=1/2$.  These similarities suggest that we may be able
to explain all these additional features using a single underlying
origin.  The asymmetry of the sashes about the Fermi energy rules out
inelastic tunnelling as a possible origin.  Their appearance in
quantum wells in several different heterostructures excludes
defect-induced resonances in the tunnelling barrier.

One hypothesis, suggested by the location of the $\nu=1$ sash, would be
the identification of the quasiparticles as ``skyrmions,'' a coupled
spin-charge excitation known to be important near $\nu=1$
\cite{Sondhi93}.  Although skyrmions exist as part of the ground state
only near $\nu=1$\cite{Barret95,Schmeller95,Gallais08}, they may exist
as high-energy excitations away from $\nu=1$ where the sash is
visible.  However, the energy and existence of skyrmions is sensitive
to the total magnetic field, not only the perpendicular magnetic
field\cite{Schmeller95}.  Measurements with tilted magnetic fields do
not reveal any change in the energy of the sash
(See suppl. figure 2).  Furthermore, the presence of a second sash
(purple arrow in \autoref{data_figure}\textbf{b}) in the fan of
features at $\nu>1$ is inconsistent with this interpretation.

\begin{figure}
\ifthenelse{\boolean{figs}}{\resizebox{\dcolwid}{!}{\includegraphics{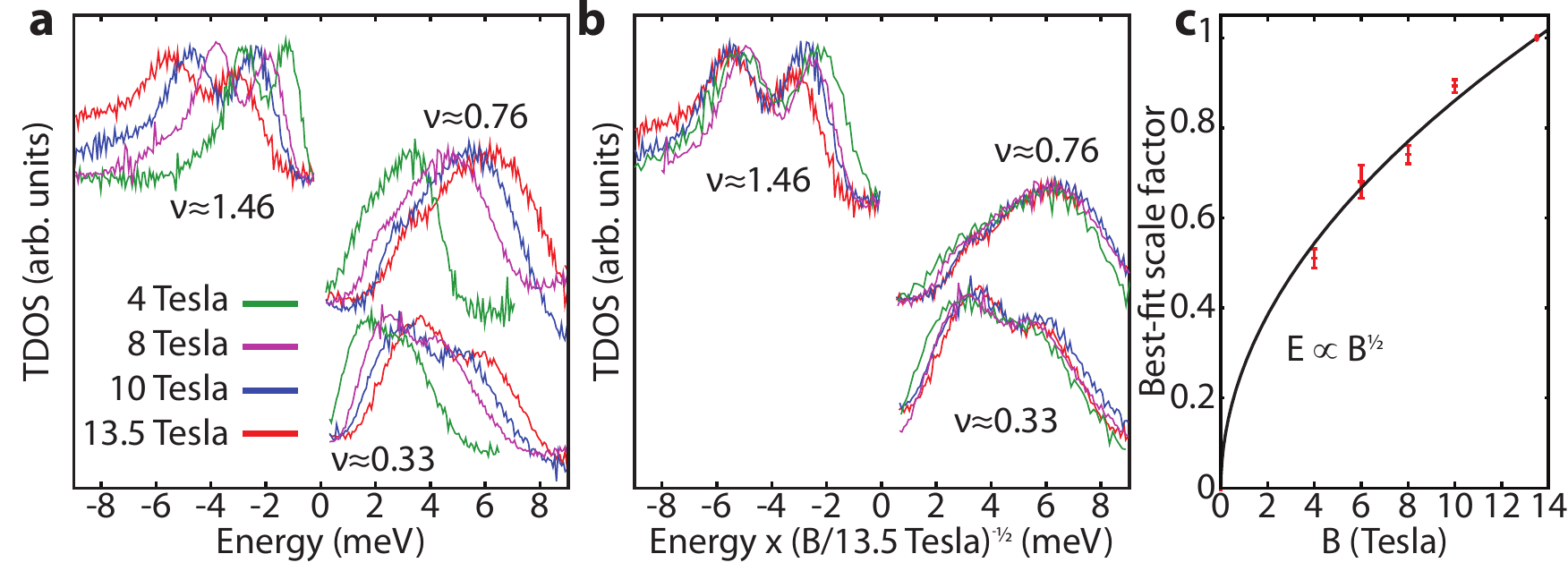}}}{}
\caption{\textbf{A $\sqrt{B}$ energy dependence indicates the sashes
  originate with electron-electron interactions within the lowest
  Landau levels}.  Linecuts at several magnetic fields, all at 100 mK,
  and at a variety of filling factors are plotted in \textbf{a},
  showing a number of peaks owing to the unexpected sashes.  In
  \textbf{b}, the energy (horizontal) axis has been scaled by
  $1/\sqrt{B}$, collapsing the peaks associated with the
  $\nu=1/2$ and $\nu=1$ sashes.  Residual
  mismatch in the peak locations is predominantly due to small
  mismatches in the filling fraction selected at each magnetic fields.
Best-fit scaling factors averaged across all densities for each magnetic field
  are shown in \textbf{c}, with a square-root dependence on
  $B$ shown for reference.  Error bars indicate the sample standard deviation
  for the best-fit scaling factor as a function of density.  For fitting
  details and alternate functional dependences see supplement.}
\label{massfig}
\end{figure}

In the absence of a model for the lineshape of these new features, it is difficult to extract their exact energies to study their dependence on magnetic field.  However, just as we can identify the linear dependence of the ``sash'' energies with density by examining their evolution with filling fraction at fixed magnetic field, we can also make precise comparisons between spectra at different $B$ but similar $\nu$ (Figure 2\textbf{a}).  We find an energy scaling factor that depends only on $B$ can be used to collapse all of the peaks in each spectrum onto the 13.5 Tesla spectrum (Figure 2\textbf{b}).  This scaling factor grows as $\sqrt{B}$ (Figure 2\textbf{c}).  The only fundamental energy scale in this system with this magnetic field dependence is the Coulomb energy scale, proportional to the inverse inter-electron separation.  This $\sqrt{B}$ scaling indicates that it may be possible to interpret the new ``sash'' features in terms of composite Fermions, weakly interacting quasiparticles whose properties are determined entirely by electron-electron interactions.

One interpretation of the paired nature of our sashes above and below
the Fermi energy is that they are part of a single, continuous
excitation that projects through the Fermi energy, with the region
near zero energy obscured by the Coulomb gap.  The $\nu=1/2$ sash then
extrapolates through $E_f$ at $\nu=1/2$ (green arrows in Figure
1\textbf{b,c,e}) Composite Fermions with two attached flux quanta
(\CFT{}s) experience an effective magnetic field $B^*=B(1-2\nu)$.  If
we imagine sweeping the density in the 2DES while injecting a probe
\CFT\, the \CFT\ will experience zero magnetic field at $\nu=1/2$,
with an effective magnetic field that increases as one moves away from
$\nu=1/2$ (red line in Figure 3a).  A Landau fan of \CFT{}s would then
follow the same trend in energy that we see in our $\nu=1/2$ sash
(Figure 3\textbf{c}).  We measure our spectra relative to the Fermi
energy, which is the same for composite fermions and
electrons\cite{Halperin93}; shifting the Landau fan to place $E_f$ at
$E=0$ yields the fan in Figure 3\textbf{d}.  Note the whole electron
may not be able to directly break up into composite Fermions; if the
CF interpretation is correct, the features in our spectrum are clouds
of incoherent excitations whose lower energy edge is the true CF
energy (see supplement).

\begin{figure}
\ifthenelse{\boolean{figs}}{\resizebox{\scolwid}{!}{\includegraphics{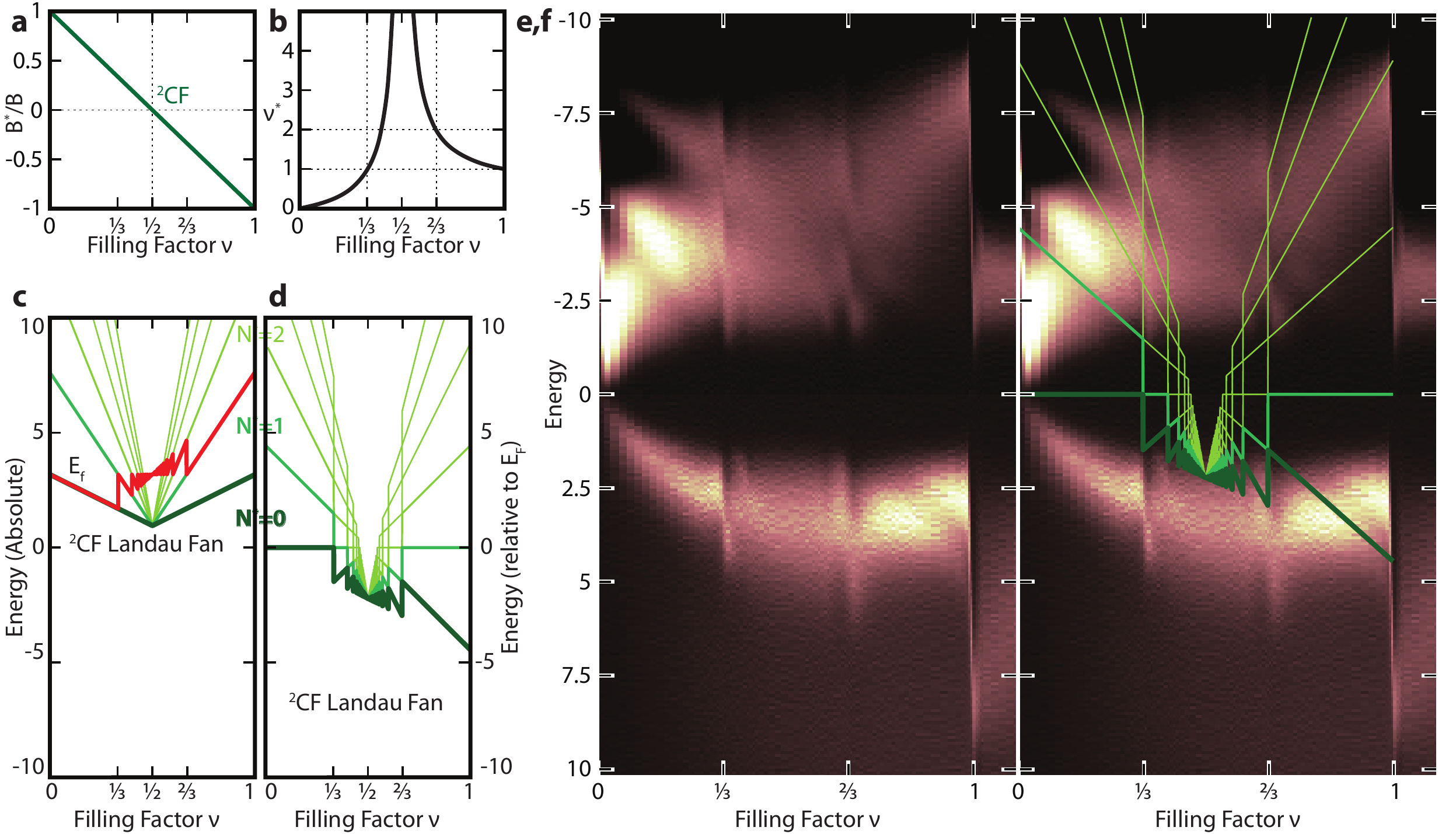}}}{}
\caption{\textbf{From the viewpoint of a composite quasiparticle,
  sweeping the density in a quantum well also sweeps the effective
  magnetic field.}  Panel \textbf{a} shows the effective field
  experienced by a \CFT\ as the electron density is swept through the
  first Landau level.  As the effective field and particle density are
  swept, the filling factor of \CFT\ composite fermions
  $\nu^*=\nu/(1-2\nu)$ changes (\textbf{b}).  In \textbf{c}, a Landau
  fan, simplified by assuming a constant CF effective mass, $E=\hbar
  \omega_c (n+1/2)$ for the \CFT\ is shown in absolute energy, with
  the Fermi energy shown in red: $E_f=\hbar \omega_c (n_f + 1/2)$,
  with $n_f$ the greatest integer less than $\nu^*$ ($n_f = \lfloor
  \nu^* \rfloor$).  The Fermi energy rises as the \CFT\ density
  increases in the quantum well.  In \textbf{d}, the fan is shifted to
  place $E_f$ at $E=0$, showing the fan as it is expected to appear in
  TDCS spectra.  The 13.5 Tesla TDCS spectrum without (\textbf{e}) and
  with (\textbf{f}) the fan superimposed allows identification of the
  two $\nu=1/2$ sash features with a Landau level of the \CFT\ fan closest 
  to the Fermi surface.  Lines suggested by the fan but not
  observed have been included to show alignment with chemical
  potential jumps associated with known FQH states, as well as to
  demonstrate the origin of the asymmetry when high energy CF Landau
  levels (lightest lines) are not observed. }
\label{fan}
\end{figure}

For $\nu<1/3$, the \CFT\ filling factor $\nu^*$ is less than 1
(\autoref{fan}\textbf{b}); this means that there are no quasi-Landau levels
below the Fermi energy, and we only expect to see states at $E>0$.
However, for $\nu>2/3$, $\nu^*$ is between one and two; the model predicts
a single quasi-Landau level below the Fermi energy.  Examination
of the fan in \autoref{fan}\textbf{e,f} clearly shows this expected
asymmetry. We find that we can observe this sash over a
wide variety of filling fractions, including those corresponding to
FQH plateaus caused by other quasiparticles.  This is consistent with
prior experiments\cite{Hirjibehedin03} and theoretical
work\cite{Jain04}.  We do not observe \CFT\ Landau levels
that would correspond to $n^* \ge 2$ in these spectra.  In the regions where we
observe \CFT\ Landau levels, these would be high energy excitations 
that should not be well described by composite Fermion theory and may
not even exist.  Unsurprisingly, we do not observe \CFT\ states
near $\nu=1/2$ where they would be expected to fall inside of the
Coulomb gap.

In this interpretation, the slopes of the lines in our Landau fans
$dE/d\nu$ relate directly to the quasiparticle effective masses; for
each \CFT\ splitting, $E = \pm \hbar e B(1-2\nu)/m^*$, so $dE/d\nu =
\pm \hbar e 2 B / m^*$.  Using this relation we estimate a mass of
$(0.35 \pm 0.06) m_0$, where $m_0$ is the free electron mass, for the
\CFT\ at 13.5 Tesla magnetic field.  Theoretical estimates of the
width of the gap at 13.5 Tesla give typical values of $m^*/m_0$ of
0.23\cite{Morf95} or 0.31 \cite{Halperin93}.  Finite well width
effects will increase these masses slightly\cite{Park99}.  The
aforementioned $\sqrt{B}$ dependence of the sash features yields a
$\sqrt{B}$ dependence of the cyclotron mass, as is predicted and as
has been observed in thermally activated transport measurements
\cite{Leadley96}.  The observed Landau gap for the $\nu=1/2$ sash is
roughly a factor of four larger than that reported from thermally
activated conductivity, and the mass is correspondingly
smaller\cite{Leadley96}. However, similar discrepancies in
measurements of the exchange gap suggest that thermal activation
measurements may underestimate the widths of gaps at the Fermi
energy\cite{Dial07}.

\begin{figure}
\ifthenelse{\boolean{figs}}{\resizebox{\scolwid}{!}{\includegraphics{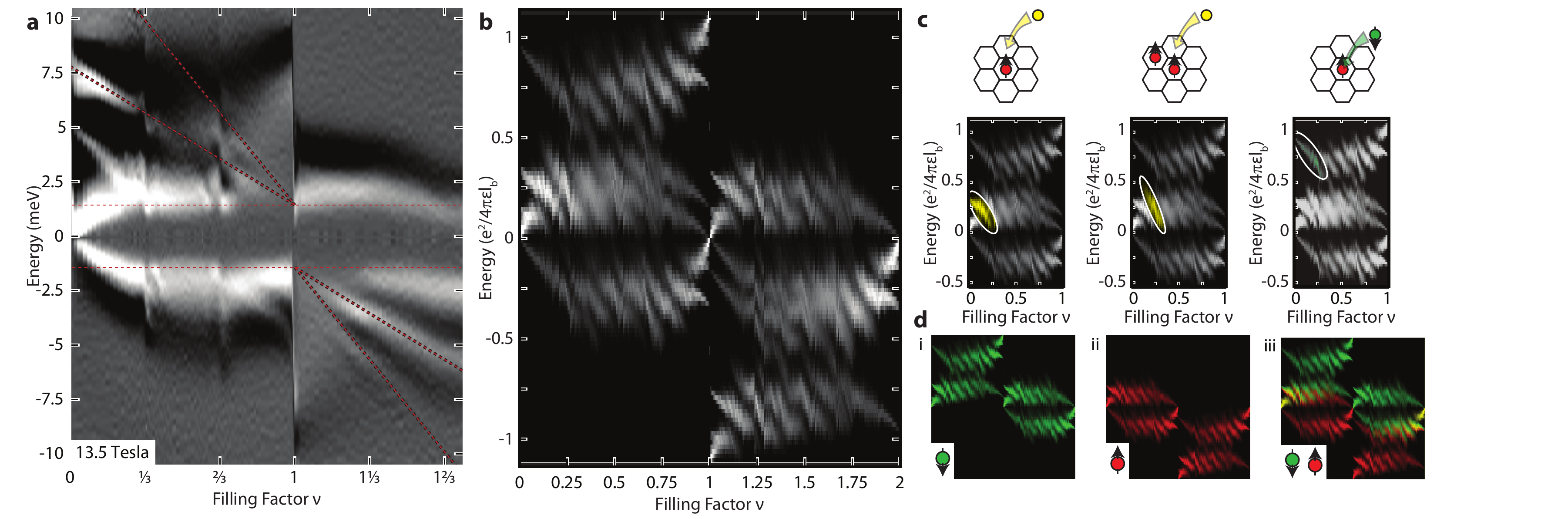}}}{}
\caption{\textbf{The $\nu=1$ sash features can be emphasized by taking an
  additional derivative of the data, easing comparison to simulated
  spectra from a model of electrons localized on a lattice.}
  Panel \textbf{a} shows the wide range of fractions over which the $\nu=1$ sash
  persists.  Noise introduced in the image by taking the second
  derivative has been diminished by smoothing with a $\sigma=160 \mu$eV
  Gaussian.  The onset energies of the ``clouds'' of excitations are
  evenly spaced in energy, allowing a simple fan to describe all of
  the excitations.  In \textbf{b}, a sample spectrum from the lattice
  model is shown.  The right axis is in units of $e^2/(4 \pi \epsilon
  l_b)$, roughly 16 meV at 13.5 Tesla.  The ``sashes'' that move
  downward in energy as the density is raised originate with different
  types of sites that can be populated or depopulated.  The origins of
  selected prominent features are indicated with cartoons of the
  relevant lattice situations in \textbf{c}.  Each cartoon is
  accompanied by a replica of the spectrum for $0<\nu<1$ with the
  relevant portion highlighted. The hexagons represent lattice cells
  of a hexagonal lattice, with single particle states in the center
  of each hexagon.  Tunneling an electron of either spin (yellow) onto
  a site adjacent to an occupied site (i) may give rise to our
  $\nu=1/2$ sash, the lowest energy sash at low densities.  A similar
  sash occurs at slightly higher densities (ii) corresponding to
  tunneling an electron onto a site adjacent to two occupied sites.
  Tunneling a minority spin (green) electron onto a site already
  occupied by a majority spin (red) electron (iii) appears to give
  rise to our $\nu=1$ sash, the highest energy sash at low densities.
  For reference, a breakdown of the spectrum by spin is provided in
  \textbf{d}, with the minority spins only (i), the majority spins
  only (ii), and the majority and minority spin states color coded.
  Note that the minority carriers in the lowest Landau level are
  always completely spin polarized in this model if the on-site
  repulsion is larger than the peak-to-peak disorder amplitude. }
\label{cf1}
\end{figure}

The $\nu=1$ sash features do not afford a simple explanation in terms
of composite Fermions; extending the model to other composite
quasiparticles with $p$ vortices attached allows us to explain any
``sashes'' which intersect the Fermi energy at a filling factors
$\nu=1/p$ where $p$ is even\cite{Jain89}.  For a fully spin polarized
excitation, $p=1$ would correspond to an excitation which has the wrong
symmetry under exchange to be an electron wavefunction.  However, the
$\nu=1$ sash features form a ``fan'' of states with evenly spaced
energies (\autoref{cf1}\textbf{a}), suggesting it may be possible to
explain them in terms of injecting electrons into states
that experience a density dependent effective magnetic field.  

The theory of composite Fermions makes an ansatz to identify weakly
interacting quasiparticles that give rise to the fractional quantum
Hall effect.  It is expected to accurately capture the low energy
physics near $\nu=1/2$.  Given that our observations are at quite
large energies and far from $\nu=1/2$, it is remarkable that this
identification appears consistent with our results.  Despite this, it
is worth noting that inelastic light scattering measurements of for
$1/5<\nu<1/3$ measure collective excitations of composite Fermions at
energies which, when adjusted for well width and magnetic field, are
comparable to our \CFT\ energies \cite{Hirjibehedin03}.

With these concerns in mind, we note the sashes persists down to very
low densities where it seems likely a semiclassical description is
possible.  One very simple model is that described by Fogler et
al\cite{Fogler96}; consider a hexagonal lattice of sites, each of
which has an occupation number for each electron spin
$n_i^{\uparrow\downarrow}$ equal to zero or one (each site can have
zero, one, or two electrons) and uncorrelated Gaussian distributed
disorder potential.  The inter-site separation is selected such that
the lattice is completely filled at $\nu=1$.  The electrons interact
through a repulsive effective potential that includes screening from
the tunnel electrode, while electrons of the same spin experience an
additional very short range attractive exchange potential.  For each
density, the total energy is minimized by rearranging the electrons
and the excited state spectrum is calculated (see supplement for
details).

In the resulting spectra (figure \autoref{cf1}\textbf{b}), we see, as
expected, the development of an exchange splitting as the total
density approaches $\nu=1$\cite{Ando74} as well as a Coloumb gap at
zero energy\cite{Pikus95}.  The calculated exchange splitting is much
larger than the observed, possibly due to the aforementioned finite
well width effects.  Several additional bands of states emerge that
are reminiscent of the sashes in our spectra.  These bands are
insensitive to the exact heterostructure: screening by the tunneling
electrode is unimportant.  Near zero density, the lattice is populated
by a small number of spin up (majority spin) electrons located at
minima in the disorder potential.  The highest energy ``sash'' here
corresponds to putting a spin down (minority spin) electron onto a
site that has already been occupied by a spin up electron; it is
displaced upwards in energy by roughly $e^2/(4 \pi \epsilon l_b)$ by
the on-site Coulomb repulsion.  The next highest energry ``sash''
corresponds to putting either spin electron onto a site that neighbors
an occupied site.  These bands have a similar structure to the
experimentally observed $\nu=1$ and $\nu=1/2$ sashes respectively.
The energies of these features decrease as the density is raised
because the Fermi energy, the energy required to add an electron to
the system after allowing the system to relax, is increasing; as the
system is filled, new electrons are forced to be placed closer to
their nearest neighbors.  Eventually, the type of site associated with
each sash will become the lowest energy type of site left to fill,
causing the sash to merge with the Coulomb gap.  For example, the sash
associated with diagram (i) in \autoref{cf1}\textbf{c} merges with the
Fermi energy at $\nu=1/4$.  The sash originating with diagram (iii) is
more complicated.  At low densities, this sash corresponds to
injecting a minority spin electron into already occupied lattice
sites.  However, near $\nu=1$, the simulated sash's physical origin
has changed, now corresponding to adding an up or down spin electron
on top of a site that has not been occupied.  This suggests that at
low densities the $\nu=1$ sash is an electrostatically unfavorable
excitation involving a spin-singlet composed of the injected electron
and an electron already present in the 2DES.  Additional features in
the simulation not seen in our measurements correspond to clusters of
more than two electrons and are probably an artifact of the lattice
onto which we have forced the electrons.

We expect this lattice model to be a fair description near $\nu=0$ and
$\nu=1$ where the 2DEG can be described as a Wigner lattice of
disorder pinned electrons (making our initial choice of localized
single particle states accurate).  The main physics the model then
brings to light is the emergence of several discrete high energy bands
due to quantization of the electron separations at short distances.
However, even in these regimes we note that all of the sashes in the
calculation have the same slope, while several distinct slopes are
present in the experimental data.  The lattice model should become
increasingly inaccurate at the intermediate densities where the
artificial localization of the electronic states will fail.  Here, the
composite Fermion model may be more appropriate.  Both models predict
similar features at the same energies and agree roughly with our
experimental spectra but also predict features that do not appear in
our spectrum.  It is unclear if these two views each have a distinct
realm of validity or if they are equivalent in some range of energies
and densities in the sense of approximately describing the same
physics.

Whatever the precise explanation for the sashes, their existence and
prominence in a fundamental spectrum of a system otherwise so
thoroughly studied comes a stunning surprise.  A detailed and accurate
explanation of this spectrum will provide key insights into our
microscopic understanding of the ground state of a 2D electronic
system.

\singlespacing
\bibliographystyle{naturemag}
\bibliography{main}

\newcommand{\noopsort}[1]{} \newcommand{\printfirst}[2]{#1}
  \newcommand{\singleletter}[1]{#1} \newcommand{\switchargs}[2]{#2#1}
\begin{thebibliography}{10}
\expandafter\ifx\csname url\endcsname\relax
  \def\url#1{\texttt{#1}}\fi
\expandafter\ifx\csname urlprefix\endcsname\relax\def\urlprefix{URL }\fi
\providecommand{\bibinfo}[2]{#2}
\providecommand{\eprint}[2][]{\url{#2}}

\bibitem{Jain}
\bibinfo{author}{Jain, J.~K.}
\newblock \emph{\bibinfo{title}{Composite fermions}}
  (\bibinfo{publisher}{Cambridge University Press}, \bibinfo{year}{2007}).

\bibitem{Chan97}
\bibinfo{author}{Chan, H.~B.}, \bibinfo{author}{Glicofridis, P.~I.},
  \bibinfo{author}{Ashoori, R.~C.} \& \bibinfo{author}{Melloch, M.~R.}
\newblock \bibinfo{title}{Universal linear density of states for tunneling into
  the two-dimensional electron gas in a magnetic field}.
\newblock \emph{\bibinfo{journal}{Phys. Rev. Lett.}}
  \textbf{\bibinfo{volume}{79}}, \bibinfo{pages}{2867--2870}
  (\bibinfo{year}{1997}).

\bibitem{Dial07}
\bibinfo{author}{Dial, O.~E.}, \bibinfo{author}{Ashoori, R.~C.},
  \bibinfo{author}{Pfeiffer, L.~N.} \& \bibinfo{author}{West, K.~W.}
\newblock \bibinfo{title}{High-resolution spectroscopy of two-dimensional
  electron systems}.
\newblock \emph{\bibinfo{journal}{Nature}} \textbf{\bibinfo{volume}{448}},
  \bibinfo{pages}{176--179} (\bibinfo{year}{2007}).

\bibitem{Luhman07}
\bibinfo{author}{Luhman, D.~R.}, \bibinfo{author}{Tsui, D.~C.},
  \bibinfo{author}{Pfeiffer, L.~N.} \& \bibinfo{author}{West, K.~W.}
\newblock \bibinfo{title}{Electronic transport studies of a systematic series
  of {GaAs/AlGaAs} quantum wells}.
\newblock \emph{\bibinfo{journal}{Appl. Phys. Lett.}}
  \textbf{\bibinfo{volume}{91}}, \bibinfo{pages}{072104}
  (\bibinfo{year}{2007}).

\bibitem{Ashoori90}
\bibinfo{author}{Ashoori, R.~C.}, \bibinfo{author}{Lebens, J.~A.},
  \bibinfo{author}{Bigelow, N.~P.} \& \bibinfo{author}{Silsbee, R.~H.}
\newblock \bibinfo{title}{Equilibrium tunneling from the 2-dimensional
  electron-gas in {GaAs} - evidence for a magnetic-field-induced energy-gap}.
\newblock \emph{\bibinfo{journal}{Phys. Rev. Lett.}}
  \textbf{\bibinfo{volume}{64}}, \bibinfo{pages}{681--684}
  (\bibinfo{year}{1990}).

\bibitem{Eisenstein92}
\bibinfo{author}{Eisenstein, J.~P.}, \bibinfo{author}{Pfeiffer, L.~N.} \&
  \bibinfo{author}{West, K.~W.}
\newblock \bibinfo{title}{Coulomb barrier to tunneling between parallel
  two-dimensional electron systems}.
\newblock \emph{\bibinfo{journal}{Phys. Rev. Lett.}}
  \textbf{\bibinfo{volume}{69}}, \bibinfo{pages}{3804--3807}
  (\bibinfo{year}{1992}).

\bibitem{Ashoori93}
\bibinfo{author}{Ashoori, R.~C.}, \bibinfo{author}{Lebens, J.~A.},
  \bibinfo{author}{Bigelow, N.~P.} \& \bibinfo{author}{Silsbee, R.~H.}
\newblock \bibinfo{title}{Energy gaps of the two-dimensional electron gas
  explored with equilibrium tunneling spectroscopy}.
\newblock \emph{\bibinfo{journal}{Phys. Rev. B}} \textbf{\bibinfo{volume}{48}},
  \bibinfo{pages}{4616--4628} (\bibinfo{year}{1993}).

\bibitem{Yang93}
\bibinfo{author}{Yang, S.-R.~E.} \& \bibinfo{author}{MacDonald, A.~H.}
\newblock \bibinfo{title}{{C}oulomb gaps in a strong magnetic field}.
\newblock \emph{\bibinfo{journal}{Phys. Rev. Lett.}}
  \textbf{\bibinfo{volume}{70}}, \bibinfo{pages}{4110--4113}
  (\bibinfo{year}{1993}).

\bibitem{Eisenstein94}
\bibinfo{author}{Eisenstein, J.~P.}, \bibinfo{author}{Pfeiffer, L.~N.} \&
  \bibinfo{author}{West, K.~W.}
\newblock \bibinfo{title}{Compressibility of the two-dimensional electron gas:
  Measurements of the zero-field exchange energy and fractional quantum {H}all
  gap}.
\newblock \emph{\bibinfo{journal}{Phys. Rev. B}} \textbf{\bibinfo{volume}{50}},
  \bibinfo{pages}{1760--1778} (\bibinfo{year}{1994}).

\bibitem{Khrapai07}
\bibinfo{author}{Khrapai, V.~S.} \emph{et~al.}
\newblock \bibinfo{title}{Direct measurements of fractional quantum {H}all
  effect gaps}.
\newblock \emph{\bibinfo{journal}{Phys. Rev. Lett.}}
  \textbf{\bibinfo{volume}{99}}, \bibinfo{pages}{086802}
  (\bibinfo{year}{2007}).

\bibitem{Sondhi93}
\bibinfo{author}{Sondhi, S.~L.}, \bibinfo{author}{Karlhede, A.},
  \bibinfo{author}{Kivelson, S.~A.} \& \bibinfo{author}{Rezayi, E.~H.}
\newblock \bibinfo{title}{Skyrmions and the crossover from the integer to
  fractional quantum {H}all effect at small {Z}eeman energies}.
\newblock \emph{\bibinfo{journal}{Phys. Rev. B}} \textbf{\bibinfo{volume}{47}},
  \bibinfo{pages}{16419--16426} (\bibinfo{year}{1993}).

\bibitem{Barret95}
\bibinfo{author}{Barrett, S.~E.}, \bibinfo{author}{Dabbagh, G.},
  \bibinfo{author}{Pfeiffer, L.~N.}, \bibinfo{author}{West, K.~W.} \&
  \bibinfo{author}{Tycko, R.}
\newblock \bibinfo{title}{Optically pumped {NMR} evidence for finite-size
  skyrmions in {GaAs} quantum wells near {L}andau level filling $\nu{}=1$}.
\newblock \emph{\bibinfo{journal}{Phys. Rev. Lett.}}
  \textbf{\bibinfo{volume}{74}}, \bibinfo{pages}{5112--5115}
  (\bibinfo{year}{1995}).

\bibitem{Schmeller95}
\bibinfo{author}{Schmeller, A.}, \bibinfo{author}{Eisenstein, J.~P.},
  \bibinfo{author}{Pfeiffer, L.~N.} \& \bibinfo{author}{West, K.~W.}
\newblock \bibinfo{title}{Evidence for skyrmions and single spin flips in the
  integer quantized {H}all effect}.
\newblock \emph{\bibinfo{journal}{Phys. Rev. Lett.}}
  \textbf{\bibinfo{volume}{75}}, \bibinfo{pages}{4290--4293}
  (\bibinfo{year}{1995}).

\bibitem{Gallais08}
\bibinfo{author}{Gallais, Y.}, \bibinfo{author}{Yan, J.},
  \bibinfo{author}{Pinczuk, A.}, \bibinfo{author}{Pfeiffer, L.~N.} \&
  \bibinfo{author}{West, K.~W.}
\newblock \bibinfo{title}{Soft spin wave near $nu = 1$: Evidence for a magnetic
  instability in skyrmion systems}.
\newblock \emph{\bibinfo{journal}{Phys. Rev. Lett.}}
  \textbf{\bibinfo{volume}{100}}, \bibinfo{pages}{086806}
  (\bibinfo{year}{2008}).

\bibitem{Halperin93}
\bibinfo{author}{Halperin, B.~I.}, \bibinfo{author}{Lee, P.~A.} \&
  \bibinfo{author}{Read, N.}
\newblock \bibinfo{title}{Theory of the half-filled {L}andau level}.
\newblock \emph{\bibinfo{journal}{Phys. Rev. B}} \textbf{\bibinfo{volume}{47}},
  \bibinfo{pages}{7312--7343} (\bibinfo{year}{1993}).

\bibitem{Hirjibehedin03}
\bibinfo{author}{Hirjibehedin, C.~F.}, \bibinfo{author}{Pinczuk, A.},
  \bibinfo{author}{Dennis, B.~S.}, \bibinfo{author}{Pfeiffer, L.~N.} \&
  \bibinfo{author}{West, K.~W.}
\newblock \bibinfo{title}{Crossover and coexistence of quasiparticle
  excitations in the fractional quantum {H}all regime at $\nu{}\le{}1/3$}.
\newblock \emph{\bibinfo{journal}{Phys. Rev. Lett.}}
  \textbf{\bibinfo{volume}{91}}, \bibinfo{pages}{186802}
  (\bibinfo{year}{2003}).

\bibitem{Jain04}
\bibinfo{author}{Peterson, M.~R.} \& \bibinfo{author}{Jain, J.~K.}
\newblock \bibinfo{title}{Flavor altering excitations of composite fermions}.
\newblock \emph{\bibinfo{journal}{Phys. Rev. Lett.}}
  \textbf{\bibinfo{volume}{93}}, \bibinfo{pages}{046402}
  (\bibinfo{year}{2004}).

\bibitem{Morf95}
\bibinfo{author}{Morf, R.} \& \bibinfo{author}{d'Ambrumenil, N.}
\newblock \bibinfo{title}{Stability and effective masses of composite fermions
  in the first and second {L}andau level}.
\newblock \emph{\bibinfo{journal}{Phys. Rev. Lett.}}
  \textbf{\bibinfo{volume}{74}}, \bibinfo{pages}{5116--5119}
  (\bibinfo{year}{1995}).

\bibitem{Park99}
\bibinfo{author}{Park, K.}, \bibinfo{author}{Meskini, N.} \&
  \bibinfo{author}{Jain, J.~K.}
\newblock \bibinfo{title}{Activation gaps for the fractional quantum {H}all
  effect: realistic treatment of transverse thickness}.
\newblock \emph{\bibinfo{journal}{Journal of Physics: Condensed Matter}}
  \textbf{\bibinfo{volume}{11}}, \bibinfo{pages}{7283--7299}
  (\bibinfo{year}{1999}).

\bibitem{Leadley96}
\bibinfo{author}{Leadley, D.~R.}, \bibinfo{author}{van~der Burgt, M.},
  \bibinfo{author}{Nicholas, R.~J.}, \bibinfo{author}{Foxon, C.~T.} \&
  \bibinfo{author}{Harris, J.~J.}
\newblock \bibinfo{title}{Pulsed-magnetic-field measurements of the
  composite-fermion effective mass}.
\newblock \emph{\bibinfo{journal}{Phys. Rev. B}} \textbf{\bibinfo{volume}{53}},
  \bibinfo{pages}{2057--2063} (\bibinfo{year}{1996}).

\bibitem{Jain89}
\bibinfo{author}{Jain, J.~K.}
\newblock \bibinfo{title}{Composite-fermion approach for the fractional quantum
  {H}all effect}.
\newblock \emph{\bibinfo{journal}{Phys. Rev. Lett.}}
  \textbf{\bibinfo{volume}{63}}, \bibinfo{pages}{199--202}
  (\bibinfo{year}{1989}).

\bibitem{Fogler96}
\bibinfo{author}{Fogler, M.~M.}, \bibinfo{author}{Koulakov, A.~A.} \&
  \bibinfo{author}{Shklovskii, B.~I.}
\newblock \bibinfo{title}{Ground state of a two-dimensional electron liquid in
  a weak magnetic field}.
\newblock \emph{\bibinfo{journal}{Phys. Rev. B}} \textbf{\bibinfo{volume}{54}},
  \bibinfo{pages}{1853--1871} (\bibinfo{year}{1996}).

\bibitem{Ando74}
\bibinfo{author}{Ando, T.} \& \bibinfo{author}{Uemura, Y.}
\newblock \bibinfo{title}{Theory of oscillatory g factor in an {MOS} inversion
  layer under strong magnetic fields}.
\newblock \emph{\bibinfo{journal}{J. Phys. Soc. Jpn.}}
  \textbf{\bibinfo{volume}{37}}, \bibinfo{pages}{1044--1052}
  (\bibinfo{year}{1974}).

\bibitem{Pikus95}
\bibinfo{author}{Pikus, F.~G.} \& \bibinfo{author}{Efros, A.~L.}
\newblock \bibinfo{title}{Coulomb gap in a two-dimensional electron gas with a
  close metallic electrode}.
\newblock \emph{\bibinfo{journal}{Phys. Rev. B}} \textbf{\bibinfo{volume}{51}},
  \bibinfo{pages}{16871--16877} (\bibinfo{year}{1995}).

\end{thebibliography}


\newcommand{\noopsort}[1]{} \newcommand{\printfirst}[2]{#1}
  \newcommand{\singleletter}[1]{#1} \newcommand{\switchargs}[2]{#2#1}
\begin{thebibliography}{1}
\expandafter\ifx\csname url\endcsname\relax
  \def\url#1{\texttt{#1}}\fi
\expandafter\ifx\csname urlprefix\endcsname\relax\def\urlprefix{URL }\fi
\providecommand{\bibinfo}[2]{#2}
\providecommand{\eprint}[2][]{\url{#2}}

\bibitem{Jain05}
\bibinfo{author}{Jain, J.~K.} \& \bibinfo{author}{Peterson, M.~R.}
\newblock \bibinfo{title}{Reconstructing the electron in a fractionalized
  quantum fluid}.
\newblock \emph{\bibinfo{journal}{Phys. Rev. Lett.}}
  \textbf{\bibinfo{volume}{94}}, \bibinfo{pages}{186808}
  (\bibinfo{year}{2005}).

\bibitem{Vignale06}
\bibinfo{author}{Vignale, G.}
\newblock \bibinfo{title}{Integral charge quasiparticles in a fractional
  quantum {H}all liquid}.
\newblock \emph{\bibinfo{journal}{Phys. Rev. B}} \textbf{\bibinfo{volume}{73}},
  \bibinfo{pages}{073306} (\bibinfo{year}{2006}).

\bibitem{Smith85}
\bibinfo{author}{Smith, T.~P.}, \bibinfo{author}{Goldberg, B.~B.},
  \bibinfo{author}{Stiles, P.~J.} \& \bibinfo{author}{Heiblum, M.}
\newblock \bibinfo{title}{Direct measurement of the density of states of a
  two-dimensional electron gas}.
\newblock \emph{\bibinfo{journal}{Phys. Rev. B}} \textbf{\bibinfo{volume}{32}},
  \bibinfo{pages}{2696--2699} (\bibinfo{year}{1985}).

\end{thebibliography}

\textbf{Supplementary Information} is linked to the online version of the paper at www.nature.com/nature.

\textbf{Acknowledgments}  We are grateful to Alan MacDonald and Yigal Meir, who independently suggested the identification of sash features with nearby localized states that formed the basis for the lattice model. We thank Patrick Lee, Leonid Levitov and Bert Halperin for helpful discussions regarding the interpretation of our results.  This work was sponsored by the Office of Science, Basic Energy Sciences, of the US Department of Energy.

\textbf{Author Contributions} O.E.D. built the apparatus and performed measurements and analysis.  R.C.A. supervised the work and performed analysis. O.E.D. and R.C.A. prepared the manuscript.  L.N.P. and K.W.W. performed the crystal growth.

\textbf{Author Information} Reprints and permissions information is available at www.nature.com/reprints.  Correspondence and requests for materials should be addressed to O.E.D. (dial@alum.mit.edu) or R.C.A. (ashoori@mit.edu).

\end{document}


\newcommand{\CFF}{\ensuremath{^4\mathrm{CF}}}
\newcommand{\CFT}{\ensuremath{^2\mathrm{CF}}}
\newcommand{\CFO}{\ensuremath{^1\mathrm{C}}}
\newlength{\scolwid}
\newlength{\dcolwid}
\setlength{\scolwid}{3.3in} 
\setlength{\dcolwid}{6.5in} 
\author {O. E. Dial\footnote{Massachusetts Institute of Technology, Cambridge, MA 02139, USA}, R. C. Ashoori\footnotemark[1], L. N. Pfeiffer\footnote{Alcatel-Lucent Bell Laboratories, Murray Hill, New Jersey 07974, USA}, K. W. West\footnotemark[2]}
\title{Composite particle in the energy spectrum of the fractional quantum Hall effect}

%
%
\textbf{\Large Supplementary Material}
\small
\setlength{\baselineskip}{3.76mm}
\section{Discharge Pulses}
Previous measurements were limited to low magnetic fields because of
the magnetic field induced coulomb gap at the Fermi energy$^{\mathrm{}3,5-8}$; vanishing
tunnel current near equilibrium caused the re-equilibration time to
grow rapidly with increasing magnetic field, making the measurement
impractical.

We have extended the technique to include a ``discharge pulse'', a
brief pulse of the opposite polarity of the pulse used to measure the
tunnel current, whose width is tuned to remove exactly the amount of
charge that tunneled in during the application of the measurement
pulse.  This rapidly returns the device to near-equilibrium.  The
tuning is performed by monitoring the tunneling current for several
tens of microseconds (this time is increased as the magnetic field is
increased) after the discharge pulse to insure that the sample is
re-equilibrated.  This time period also allows the 2DES to
re-thermalize.  We confirm that this re-thermalization is complete by
varying the delay between pulse repetitions and confirming that we
measure the same I/V curve.

\section{Energy Scaling}
The energy scaling factor for each magnetic field shown in Figure 2 of
the main text was determined by
maximizing the correlation between $d^2 I/d V^2$ spectra taken at the
same density but at two different magnetic fields.  First, for each
filling in the higher field spectrum, the closest filling factor in the
lower field spectrum is selected.  Then, the spectra are numerically
differentiated a second time, giving $d^2 I/ d V^2$; this minimizes
any overall background to the spectrum.  High frequency noise
resulting from this double-differentiation is then removed by
convolving each spectrum with a $\sigma=0.38$ meV Gaussian.  The lower
field spectrum $\rho_1(E)$ is then stretched along the energy axis by
a scale factor $\epsilon$ and resampled to the same energy resolution
as the higher field spectrum $\rho_2(E)$ using linear interpolation.
The value of $\epsilon$ that maximizes the correlation $\chi(\epsilon)
= \int \rho_1(\epsilon E) \rho_2(E) dE$ between the two spectra is
then searched for and located.  At this point, the best estimate for
$\epsilon$ is noisy due to discretization of the data in energy.  To
remove this noise, $\chi(\epsilon)$ is evaluated at 25 values of
$\epsilon$ over a range stretching from 10\% below to 10\% above this
absolute best correlation, and the resulting curve of $\chi$ versus
$\epsilon$ is fit using a quadratic.  The value of $\epsilon$ at the
peak of this quadratic curve is then selected as the best scale factor
for aligning the two spectra at this density.

\begin{figure}
\ifthenelse{\boolean{figs}}{{\resizebox{\scolwid}{!}{\includegraphics{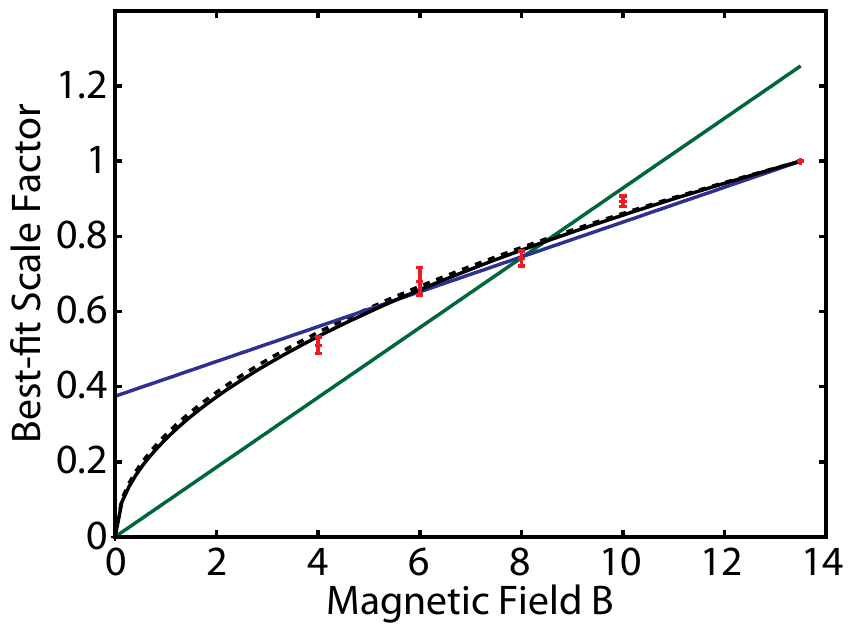}}}}{}
\caption{\textbf{Alternate functional forms for the fit to the magnetic field dependence of the scale factor are less satisfactory than $\sqrt{B}$.} Error bars indicate the standard deviation of the scale factor $\epsilon$ across the range of available densities at each magnetic field.  The fitting functions are $\sqrt{B}$ (solid black line, $\chi^2=2.4$), power-law $B^\alpha$ (dotted black line, $\alpha=.52$, $\chi^2=3.1$), linear with an offset (blue line, $\chi^2=6.9$), 
and linear without an offset (green line, $\chi^2=20$). 
\label{fitfig}}
\end{figure}

This process is repeated for every density from $0.1<\nu<0.9$ and
$1.1<\nu<1.9$ in the spectrum.  The resulting map of $\epsilon$ versus
$\nu$ is roughly constant; maximum deviations from the mean value are
of the order 10\%.  This flatness indicates that a single value of
$\epsilon$ is sufficient at all densities to explain the magnetic field
dependence of our data.  This value is found by averaging the value of
$\epsilon$ across this range of $\nu$, discarding any outliers that
fall more than 3 standard deviations away from the mean.  This mean
$\epsilon$ is used in figure 2\textbf{c}.  The standard deviation of
$\epsilon$ across this range of densities is taken as width of the
error bars, essentially the range of $\epsilon$ found to be a
reasonable fit.  As such they are not representative of normally
distributed statistical errors.  However, the reduced sum-squared
error ($\chi^2$) still a rough guide as to the quality of different
fits.  \autoref{fitfig} shows alternate fits to the B-field
dependence.

\section{Supplemental Data}
Supplementary figure 1 demonstrates that raising the sample
temperature to 4.1 Kelvin eliminates the ``sash'' features
while leaving the exchange splitting intact.  The broadening of the
Fermi function in the tunnel electrode is not sufficient to explain
this change.  Supplementary Figure 2
demonstrates the insensitivity of the $\nu=1$ sash to presence of
magnetic field in the plane. Supplementary Figure 3 displays the
effect of disorder on both the $\nu=1/2$ sash and the $\nu=1$ sash. The
$\nu=1/2$ sash is destroyed by disorder in a narrower quantum well whereas
the $\nu=1$ sash survives this disorder.

\begin{figure*}
\ifthenelse{\boolean{figs}}{{\resizebox{\dcolwid}{!}{\includegraphics{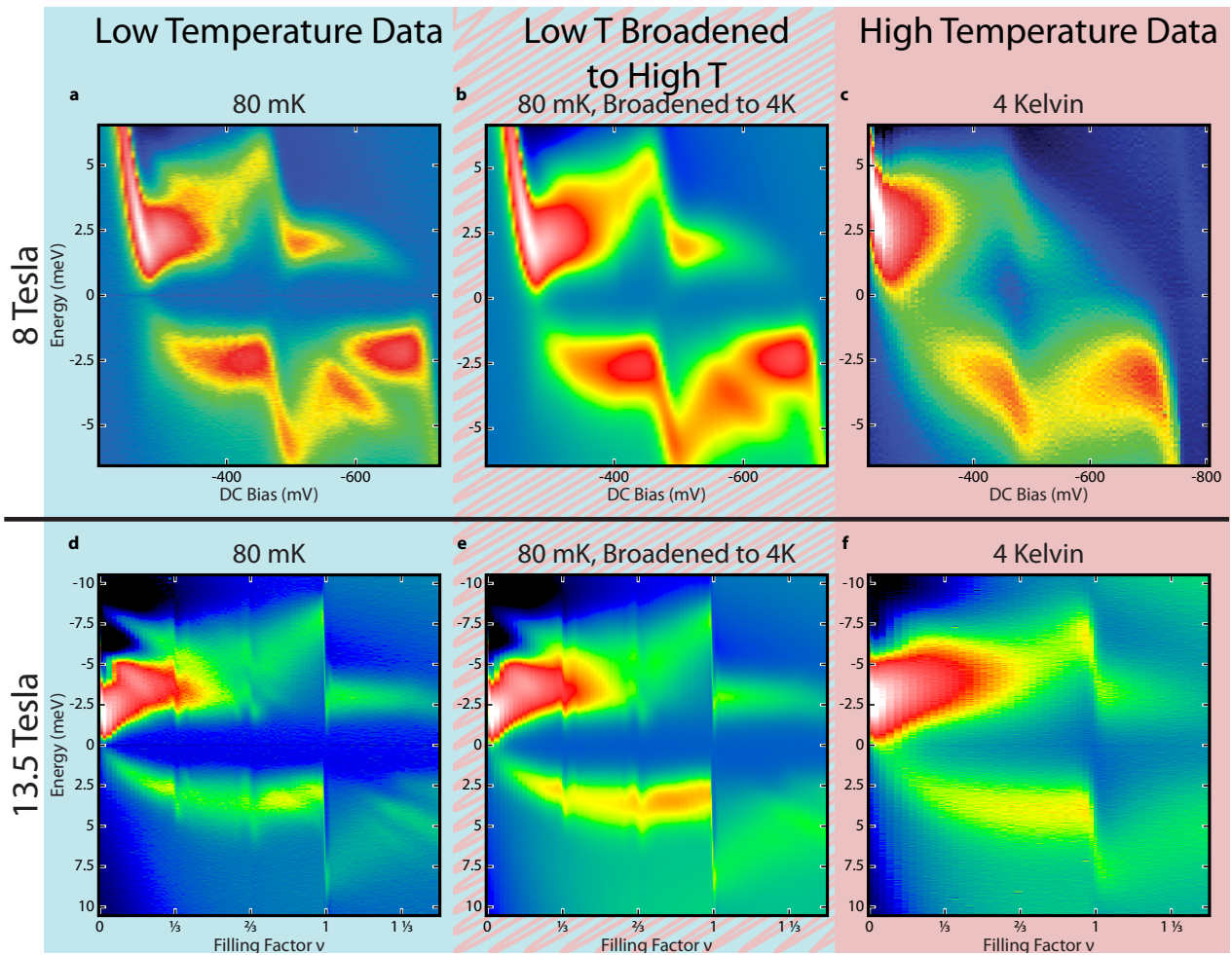}}}}{}
\caption{\textbf{\boldmath{}The effect of temperature on the $\nu=1$
    and $\nu=1/2${ }sashes} is shown by comparing a spectrum acquired
  in a 175 \AA\ quantum well at 8 Tesla and 80 mK (\textbf{a}) to one
  measured at 4.1 K (\textbf{c}).  The effect of Fermi function
  broadening in the electrode and quantum well can be demonstrated by
  convolving with a 4.1 K Fermi function (\textbf{b}).  This does not
  destroy the sash features, demonstrating that some change in the
  many-body state of the quantum well is responsible.  Similar data at
  13.5 Tesla in a 230 \AA\ well also shows the $\nu=1/2$ sash features
  completely destroyed at 4.1 Kelvin, while some very slight trace of
  the $\nu=1$ sash survives (\textbf{d-f}).
\label{temp_figure}}
\end{figure*}
\begin{figure}
\ifthenelse{\boolean{figs}}{{\resizebox{\scolwid}{!}{\includegraphics{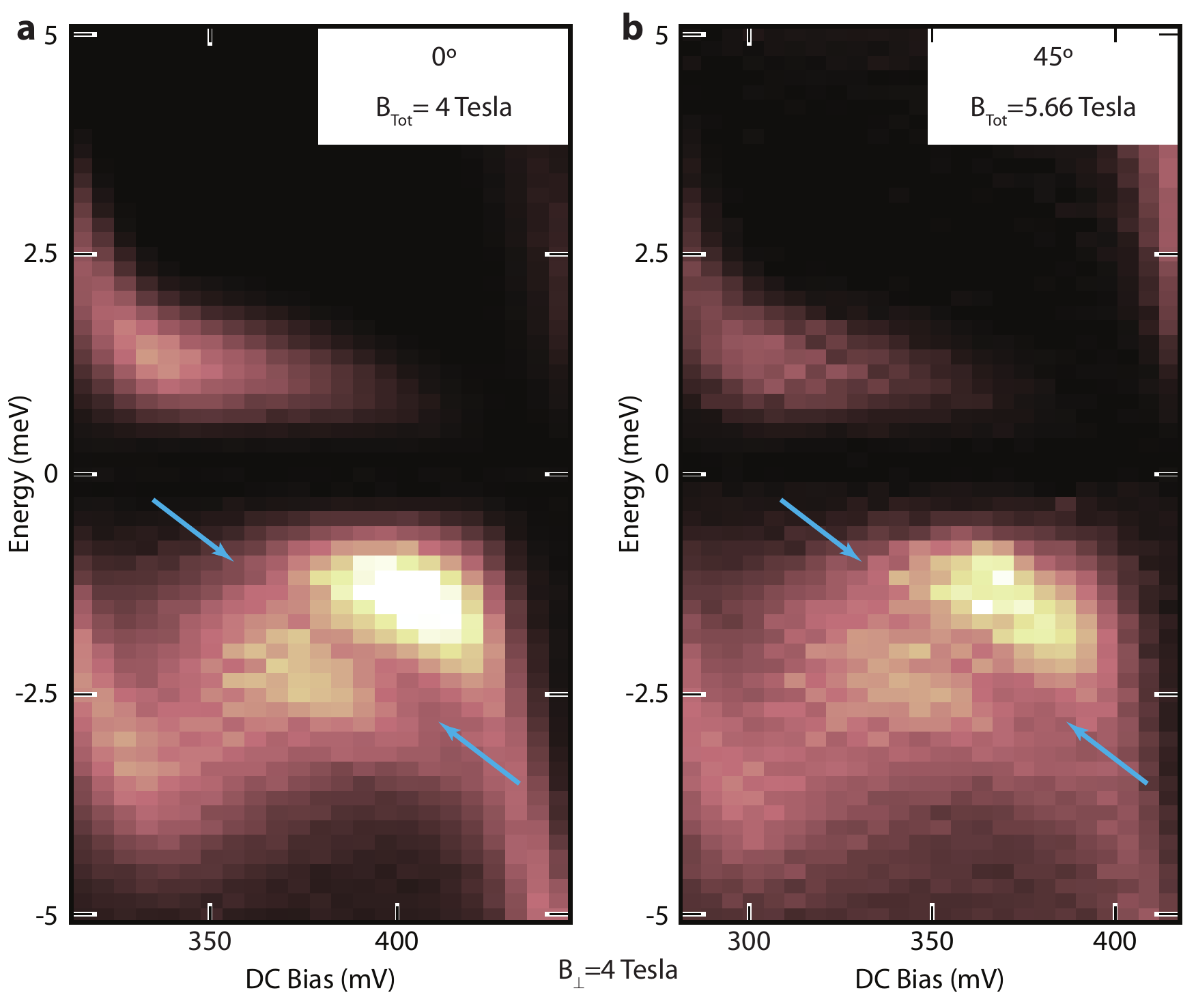}}}}{}
\caption{\textbf{Tilting the sample while maintaining the same
    magnetic field perpendicular to the quantum well} increases the
  total magnetic field applied.  This increases the Zeeman energy
  (which varies with total field), while keeping the exchange energy
  (which varies with perpendicular field) fixed.  Excitations which
  originate with the interplay between the Zeeman and exchange
  energies, including skyrmions, are sensitive to this variation of the total magnetic
  field.  By comparing the spectrum in \textbf{b} to \textbf{a}, we
  see that increasing the total magnetic field by 41\% does not affect
  the $\nu=1$ sash (blue arrows).  This data was taken in a device with
  a 175 \AA\ quantum well at 80 mK. }
\label{tilt_figure}
\end{figure}

\begin{figure}
\ifthenelse{\boolean{figs}}{{\resizebox{\scolwid}{!}{\includegraphics{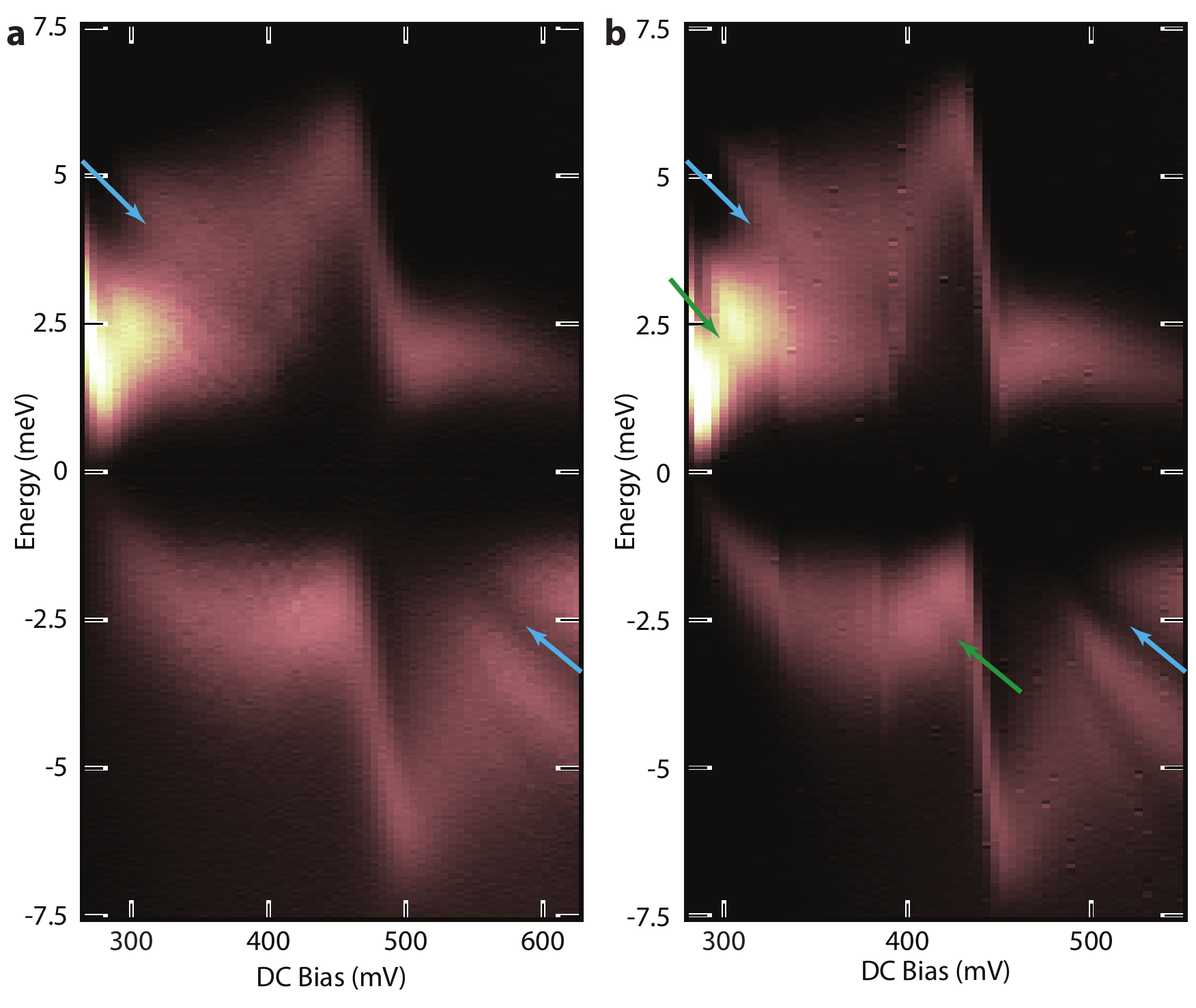}}}}{}
\caption{\textbf{\boldmath{}The effect of disorder on the $\nu=1$ and $\nu=1/2$ sashes} is shown
  by comparing a spectrum acquired in a 175 \AA\ quantum well
  (\textbf{a}) and a wider 230 \AA\ quantum well (\textbf{b}), both
  with an applied field of 4 Tesla at 80 mK.  Although
  the $\nu=1$ sash is clearly visible in both spectra (blue arrows), the
  added disorder from the narrower quantum well almost completely
  destroys the $\nu=1/2$ sash (green arrows).}
\label{disorder_figure}
\end{figure}

\section{Orthogonality Issues in the Composite Fermion Model}

Given that whole electrons are tunnelling into the system, not
composite particles, why do we observe associated structures in our
spectra?  Jain has suggested that when the 2DES is in an incompressible
fractional state, it may be possible to construct several
composite fermions from a whole electron, resulting in a peak in the
SPDOS\cite{Jain05,Vignale06}.  Our peaks sharpen as the 2DES
enters fractional Hall plateaus, which we may understand as due to an
increase in quasiparticle lifetimes as the system becomes gapped.
However, we do not observe any qualitative change in the spectra, nor
do we detect a new, sharp resonance.  The spectral weight of the peak
suggested by Jain may be too small to observe in our otherwise crowded
spectra.

Instead, we expect tunnelling directly into a CF Landau level to be
strongly suppressed.  This suppression reflects the same physics as
that responsible for the Coulomb gap at zero energy.  In the case of
the Coulomb gap, although there is a high thermodynamic density of
states at the Fermi energy\cite{Smith85} in a partially filled Landau
level, the tunnelling density of states at low energies is reduced by
inter-electron repulsion between the tunnelling electron and the
electrons already in the quantum well.  Similarly, we expect a Coulomb
pseudo-gap for tunnelling an electron into a highly correlated CF
Landau level.  The true energy of the CF level occurs at an energy
closer to $E_f$ than the ``clouds'' of incoherent excitations we
observe as the bright part of the ``sash''; this is reflected in the
placement of the arrows and fans at the edge of the bright part of the
sash in the spectra.  However, regardless of the exact relationship
between the energy of the quasiparticle and the energy of this
``cloud'', the behaviour of the ``cloud'' can be used to track the
evolution of the quasiparticle as the 2DES density and magnetic field
are varied.  This measurement is insensitive to quasiparticle charge;
when tunnelling an electron into an FQH state with fractional charge
$1/q$, conservation of charge demands that we create $q$ excitations,
each with energy $\frac{\hbar e B^*}{q m^*}$, requiring a total energy
of $\hbar e B^*/m^*$.

\section{Lattice Model}
\begin{figure}
\ifthenelse{\boolean{figs}}{{\resizebox{\scolwid}{!}{\includegraphics{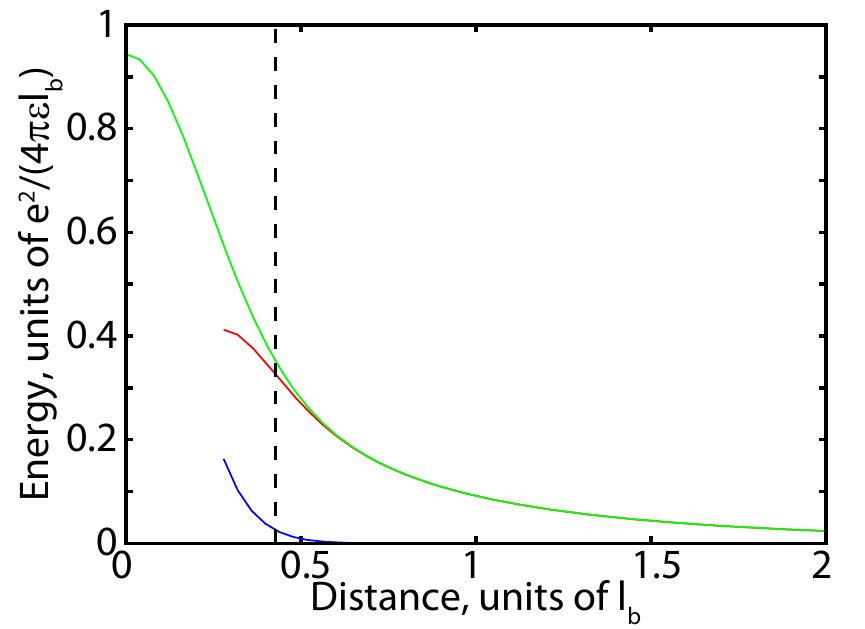}}}}{}
\caption{\textbf{Sample effective potentials calculated for the
    lattice model} show the relatively short range of the exchange
  attraction.  The repulsion between electrons of unlike spins is
  given by the green curve, while that for like spins by the red
  curve.  The blue line shows the strength of the exchange coupling
  (the difference between the red and green curves.)  Note that the
  potential for like spins is meaningless at very short distances due
  to Pauli exclusion.
  The lattice constant of the hexagonal lattice is shown with the
  black dotted line.}
\label{potential_figure}
\end{figure}

\begin{figure*}
\ifthenelse{\boolean{figs}}{{\resizebox{\dcolwid}{!}{\includegraphics{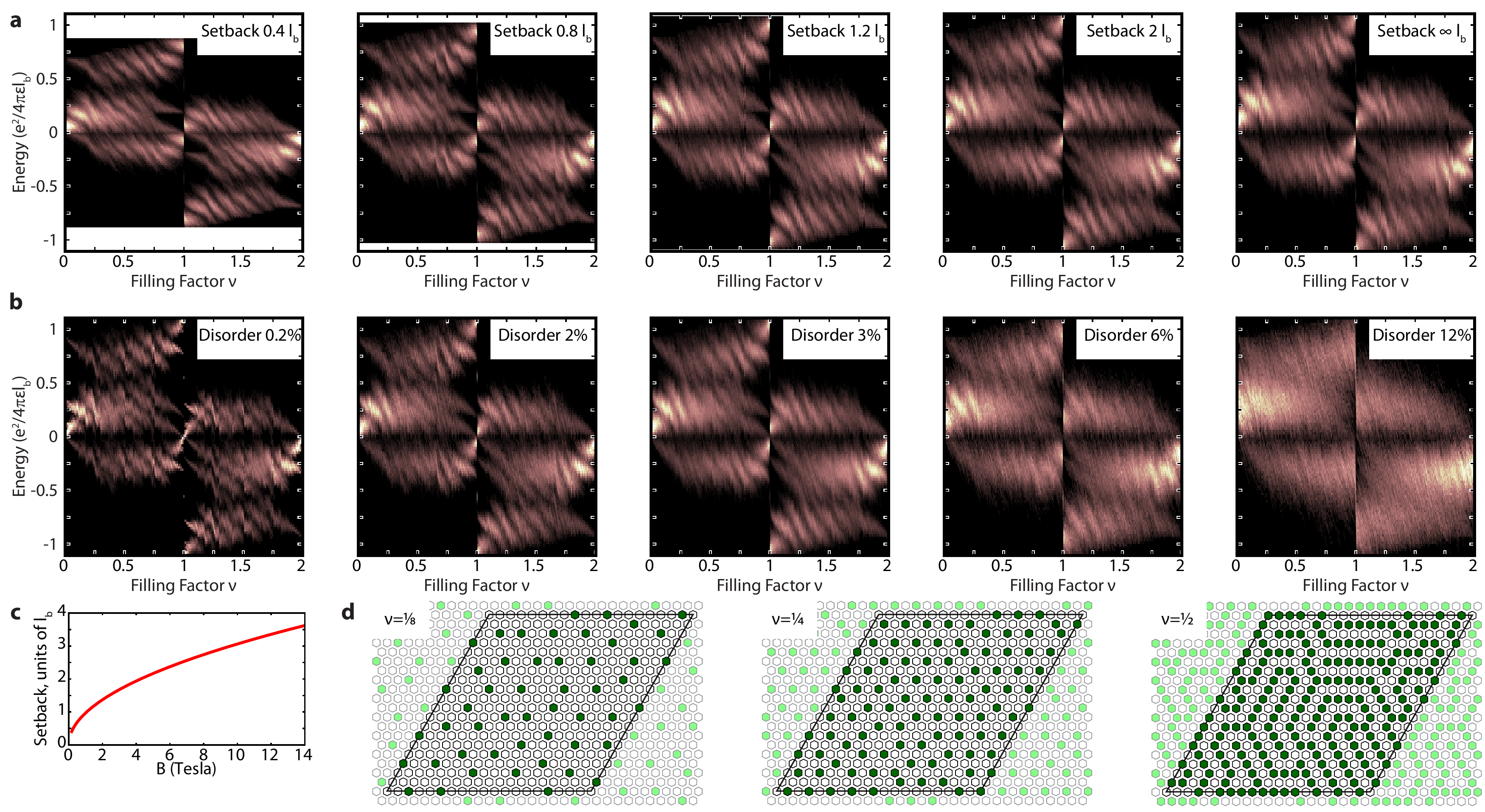}}}}{}
\caption{\textbf{Sample spectra generated by the lattice model} exhibit
  features similar to the experimentally observed sashes.
  \textbf{a} shows a series of simulations with differing setbacks
  (distances from the center of the quantum well to the 3D electrode)
  and fixed disorder of 3\% of the Coulomb energy scale
  $e^2/(4\pi\epsilon l_b$.  Note that setbacks of more than $l_b$ have
  very little impact on the spectrum.  The effects of disorder are
  explored in \textbf{b}, where a series of spectra with a fixed
  setback of 2 $l_b$ but varying disorder, measured as a percentage of
  the Coulomb energy scale, are shown.  The actual setback varies with
  magnetic field and sample design; the setback appropriate for the
  device used in these spectra is shown in \textbf{c}.  Prominent sash
  features are visible from about 4 to 13.5 Tesla, corresponding to
  setbacks of 1 to 3 $l_b$.  All spectra are calculated with a 25x25
  unit cell.  Sample annealed electron lattices are illustrated using
  a smaller 20x20 unit cell at several filling factors in \textbf{d},
  with a single unit cell shown in bold.  Note the uniformity in the
  number of occupied neighbor sites.  These lattices are for a
  disorder of 3\% and a setback of 2 $l_b$. See also supplemental
  movie A. }
\label{sim_figure}
\end{figure*}

We use a lattice model based roughly on the one described by Fogler
et. al$^{\mathrm{}22}$.  Each electron is assumed to be localized in a
coherent state on a hexagonal lattice.  The lattice constant is
selected such that one magnetic flux quantum penetrates each unit
cell.  Periodic boundary conditions are used with a hexagonal lattice,
as shown in \autoref{sim_figure}\textbf{d}.  The resulting basis set
is almost orthogonal and has the same degeneracy as the true single
particle states.  Wavefunction overlap is neglected.  Electrons of the
same spin interact through an effective potential as given by equation
60 of reference 4.  However, to address concerns that the spectrum
may be modified by the nearby tunneling
electrode, the bare Coulomb interaction $v(r)$ is replaced by $\left(
e^2/{4\pi\epsilon} \right) \left(r^{-1}-(r^2+s^2)^{-1/2}\right)$ where
$s$ is the distance between the 2D system and the nearby 3D electrode.
Electrons of different spins interact through the same potential, but
with the exchange term $u_{EX}$ set to zero
(\autoref{potential_figure}).  On each iteration of the software, a single
electron is added to the lattice at the most energetically favorable
position, and then the electron occupations are selected to minimize
the total energy as described in reference 4.


The system is then modified by adding a single electron or hole on one
lattice site, and the energy change is calculated and stored.  This
process is repeated for each possible lattice site, carrier type, and
spin.  The chemical potential is taken as the mean of the largest hole
addition energy and the smallest electron addition energy.  The energy
difference between the chemical potential and the addition energy is
then histogrammed.  This histogram is used as an approximation to the SPDOS
at this density.

The entire process is then repeated until the lattice is filled,
generating a complete simulated TDCS spectrum.  The process of filling
the lattice is demonstrated in the supplemental movie.  Sample results are
shown in \autoref{sim_figure}.  It is worth noting that in the
simulations, the nearby electrode does not play a large role once it is
more than a magnetic length away.  This condition is met in all of the
measurements described in this letter.  The main role of the screening from
this electrode is to diminish distant electron-electron interactions,
but the physics of the spectrum is dominated by nearest-neighbor
interactions (\autoref{sim_figure}\textbf{a},\textbf{c}).  

The calculated spectra show many ``sashes'' not present in the
experimental spectrum.  Although the additional sashes appear to
be artifacts of the lattice in our semi-classical model, it is
interesting to note that while the addition of disorder diminishes
all of the features in the spectrum, these additional sashes appear to
diminish more as disorder is increased (see, for example the 6\%
plot in \autoref{sim_figure}\textbf{b}).
Note that the simulation does not included any lifetime effects.  It
is not possible to estimate the disorder in our devices from these
simulations as it is likely that lifetime broadening influences
line-widths in our spectrum and which features are visible in
our spectrum.
\singlespacing
\bibliographystyle{naturemag}
\bibliography{main}